\documentclass[10pt,a4paper]{article}                                  

\usepackage[table]{xcolor}
\usepackage{algorithm,algorithmic}
\usepackage{bm}
\usepackage{mathtools}
\usepackage{graphicx}
\usepackage{cite}
\usepackage{amsmath}
\usepackage{amsfonts}
\usepackage{amssymb}
\usepackage{epstopdf}
\usepackage{subfigure}
\usepackage{booktabs}
\usepackage{rotating}
\usepackage{mathptmx}
\usepackage{color}

\DeclarePairedDelimiter\ceil{\lceil}{\rceil}
\DeclarePairedDelimiter\floor{\lfloor}{\rfloor}

\graphicspath{ {figures/} }

\begin{document}

\newtheorem{lemma}{Lemma}
\newtheorem{assumption}{Assumption}
\newtheorem{theorem}{Theorem}
\newtheorem{corollary}{Corollary}
\newtheorem{remark}{Remark}
\newtheorem{proposition}{Proposition}
\newtheorem{definition}{Definition}
\newtheorem{algo}{Algorithm}
\newtheorem{proof}{Proof}

	\date{}

	\title{On shrinking horizon move-blocking predictive control}

	\author{Hafsa Farooqi, Lorenzo Fagiano, Patrizio Colaneri\thanks{The authors are with the Dipartimento di Elettronica, Informazione e Bioingegneria, Politecnico di Milano, Milano, Italy. E-mail addresses:
			{\tt\small \{hafsa.farooqi| lorenzo.fagiano|lorenzo.fagiano| \}@polimi.it }}}

\maketitle

\section{Introduction}\label{S:intro}
This manuscript contains technical details of recent results developed by the authors on shrinking horizon predictive control with a move-blocking strategy.

\section{Motivating application: energy efficient operation of trains}\label{S:motivation_train}

Our theoretical developments are motivated by a real-world application under study in collaboration with a rail transport manufacturer, pertaining to the energy efficient operation of trains. Consider an electric train controlled by a digital control unit in discrete time, with sampling period $T_s$. Let us denote with $k\in\mathbb{Z}$ the discrete time variable, with $x(k)=[x_1(k),\,x_2(k)]^T$ the state of the train, where $x_1$ is its position and $x_2$ its speed ($\cdot^T$ denotes the matrix transpose operator), and with $u(k)\in[-1,1]$ a normalized traction force, where $u(k)=1$ corresponds to the maximum applicable traction and $u(k)=-1$ to the maximum braking. The input $u$ is the available control variable. The train has to move from one station with position $x_1=0$ to the next one, with position $x_1=x_f$, in a prescribed time $t_f$. For a given pair of initial and final stations, the track features (slopes, curvature) are known in advance. Thus, in nominal conditions (i.e. with rated values of the train parameters, like its mass and the specifications of the powertrain and braking systems), according to Newton's laws and using the forward Euler discretization method, the equations of motion of a reasonably accurate model of this system read:
\begin{equation} \label{eq:train_dynamics}
\begin{array}{l}
x_1(k+1)= x_1(k)+T_s x_2(k)\\
x_2(k+1)= x_2(k)+T_s \left(\frac{F_T(x(k),u(k))-F_B(x(k),u(k))-F_R(x(k))}{M}\right)
\end{array}
\end{equation}
\noindent where $M$ is the total mass of the train, $F_T$ is the traction force, $F_B$ is the braking force, and $F_R$ the resistive force. Functions $F_T(x,u)$, $F_B(x,u)$ are nonlinear and they depend on the specific train and track profile. They include, for example, look-up tables that link the traction and braking forces to the train speed and to the control input value. These functions are derived either experimentally or from complex models of the train and its traction and braking systems. In our research, these are provided by the business unit at our industrial partner. More details on these functions are omitted for confidentiality reasons. The resistive force $F_R(x)$ is also nonlinear, and it is the sum of a first term  $R_v(x_2)$, accounting for resistance due to the velocity, and a second term $R_g(x_1)$, accounting for the effects of slopes and track curvature:
\begin{equation} \label{eq:resistive_force}
\begin{aligned}
F_R(x)&=R_v(x_2)+R_g(x_1)\\
R_v(x_2)&= A+ Bx_2 + Cx_2^2 \\
R_g(x_1) &= M_s\left(g\tan(\alpha(x_1)) + \frac{D}{r_{c}(x_1)}\right)
\end{aligned}
\end{equation}
where the parameters $A,B,C,D$ are specific to the considered train, $M_s$ is the static mass of the train, i.e. the mass calculated without taking into account the effective inertia of the rotating components, $r_c(x_1)$ and $\alpha(x_1)$ are, respectively, the track curvature and slope at position $x_1$, and $g$ is the gravity acceleration. For example, an uphill track section corresponds to $\alpha(x_1)>0$, i.e. a positive slope.\\
Besides the prescribed arrival time $t_f$ and position $x_f$, there are additional state constraints that must be satisfied. These pertain to the limit on the maximum allowed velocity, $\overline{x}_2(x_1)$, which depends on the position $x_1$, since a different velocity limit is imposed for safety by the regulating authority according to the track features at each position. Overall, by defining the terminal time step $k_f\doteq \floor*{t_f/T_s}$ (where $\floor{\cdot}$ denotes the flooring operation to the closest integer), the state constraints read: 
\begin{equation} \label{eq:train_state_constr}
\begin{array}{ll}
x(0) &= [0,\,0]^T\\
x(k_f)&= [x_f,\,0]^T\\
x_2(k)&\geq 0,\,\,k=0,\ldots,k_f\\
x_2(k)&\leq \overline{x}_2(x_1(k)),\,\,k=0,\ldots,k_f
\end{array}
\end{equation}


The control objective is to maximize the energy efficiency of the train while satisfying the constraints above. To translate this goal in mathematical terms, different possible cost functions can be considered. In our case, we consider the discretized integral of the absolute value of the traction power over time (with a constant scaling factor $T_s^{-1}$):
\begin{equation} \label{eq:train_cost}
J =\sum\limits_{k=0}^{k_f} \left|F_T(x(k),u(k))x_2(k)\right|.
\end{equation}
This choice tends to produce controllers that minimize the traction energy injected into the system. The braking energy is not penalized, since in our case there is no restriction to the use of the braking system. \\
As already pointed out, the input variable is also constrained in the interval $u\in[-1,1]$. When the controller operates fully autonomously, i.e. without a human driver in the loop, the whole  interval can be used. However, in a driver assistance scenario, i.e. when the control algorithm is developed to assist a human driver with a suggested value of the input handle, only a smaller set of possible values can be delivered by the controller, in order to facilitate the human-machine interaction.  In particular, in this scenario the input constraints are further tightened according to four possible operating modes prescribed by our industrial partner:
\begin{itemize}
	\item \textit{Acceleration:} in this mode, the input can take one of three allowed values, i.e. $ u \in \{0.5, 0.75,1\}$.
	
	\item \textit{Coasting:} this mode implies that the traction is zero, i.e $ u=0$.
	
	\item \textit{Cruising:} in this mode, the train engages a cruise control system that keeps a constant speed, i.e. $u$ is computed by an inner control loop in such a way that $F_T = F_R$ for positive slopes and $ F_B = F_R$ for negative slopes.
	
	\item \textit{Braking:} in this mode the maximum braking force is used, i.e. $ u=-1$.
	
\end{itemize}
\noindent As a matter of fact, the modes above can be merged in just two: one with a finite integer number of possible input values $ u \in \{-1,0,0.5, 0.75,1\}$ (which unites the Acceleration, Coasting and Braking modes), and one with the cruise control engaged. Finally, a further feature of this application is a relatively small sampling time $T_s$ with respect to the imposed overall time horizon $t_f$, resulting in a rather large number of sampling periods in the interval $[0,\,t_f]$, typically from several hundreds to a few thousands.

\section{Problem abstraction and nominal SBPC approach}\label{S:nominal_MPC}
The control problem described in Section \ref{S:motivation_train} can be cast in a rather standard form:
\begin{subequations}\label{eq:FHOCP} 
	\begin{eqnarray}
\hspace{2cm}\,&\min\limits_{\bm{u}}\;\sum\limits_{k=0}^{k_f} \ell(x(k),u(k))&\\
&\text{subject to}&\nonumber\\                                
&x(k+1) = f(x(k),u(k))&\\
&u(k) \in U,\,k=0,\ldots,k_f-1&\\
&x(k) \in X,\,k=1,\ldots,k_f&\\
&x(0) = x_0&\\
&x(k_f)\in X_f&\label{eq:FHOCP_terminal_constr}
\end{eqnarray}
\end{subequations}
where $x \in \mathbb{X}\subset\mathbb{R}^n$ is the system state, $x_0$ is the initial condition, $u \in \mathbb{U}\subset\mathbb{R}^m$ is the input, $f(x,u):\mathbb{X}\times\mathbb{U}\rightarrow\mathbb{X}$ is a known nonlinear mapping representing the discrete-time system dynamics, and $l(x,u):\mathbb{X}\times\mathbb{U}\rightarrow\mathbb{R}$ is a stage cost function defined by the designer according to the control objective. The symbol $\bm{u}=\{u(0),\ldots,u(k_f-1)\}\in\mathbb{R}^{m\,k_f}$ represents the sequence of current and future control moves to be applied to the plant. The sets $X\subset\mathbb{X}$ and $U\subset\mathbb{U}$ represent the state and input constraints, and the set $X_f\subset\mathbb{X}$ the terminal state constraints, which include a terminal equality constraint as a special case. \\
We recall that a continuous function $a:\mathbb{R}^+\rightarrow\mathbb{R}^+$ is a $\mathcal{K}$-function ($a\in\mathcal{K}$) if it is strictly increasing and $a(0)=0$. Throughout this paper, we consider the following continuity assumption on the system model $f$.
\begin{assumption}\label{Ass:continuity_f}
	The function $f$ enjoys the following continuity properties:
	\begin{equation}\label{eq:continuity_f} 
		\begin{array}{l}
		\|f(x^1,u)-f(x^2,u)\|\leq a_x\left(\|x^1-x^2\|\right),\,\forall x^1,x^2\in\mathbb{X}, u\in\mathbb{U}\\
		\|f(x,u^1)-f(x,u^2)\|\leq a_u\left(\|u^1-u^2\|\right),\,\forall u^1,u^2\in\mathbb{U}, x\in\mathbb{X}
		\end{array}
	\end{equation}
	where $a_x,\,a_u\in\mathcal{K}$.\hfill$\square$
\end{assumption}
In \eqref{eq:continuity_f} and in the remainder of this paper, any vector norm $\|\cdot\|$ can be considered. Assumption \eqref{Ass:continuity_f} is  reasonable in most real-world applications, and it holds in the railway application considered here.\\
The nonlinear program \eqref{eq:FHOCP} is a Finite Horizon Optimal Control Problem (FHOCP). In the literature, many different solutions to solve this kind of a problem can be found, depending on the actual form of the system dynamics and constraints. One approach is to compute a (typically local) optimal sequence of inputs $\bm{u}^*$ and to apply it in open loop. This might be convenient when little uncertainty is present and the system is open-loop stable. Unfortunately, this is seldom the case. A much more robust approach is to resort to a feedback control policy $u(k)=\kappa(x(k))$. However, to derive explicitly such a feedback policy in closed form is generally not computationally tractable, due to the presence of system nonlinearities and constraints. A common way to derive implicitly a feedback controller is to adopt a receding horizon strategy, where the input sequence is re-optimized at each sampling time $k$ and only the first element  of such a sequence, $u^*(k)$, is applied to the plant. Then, the feedback controller is implicitly defined by the solution of a FHOCP at each $k$, where the current measured (or estimated) state $x(k)$ is used as initial condition. This approach is well-known as Nonlinear Model Predictive Control (NMPC), and is adopted here as well, however with two particular differences with respect to the standard formulation: 
\begin{itemize}
	\item First, since in our problem the terminal time $k_f$ is fixed, the resulting strategy features a shrinking horizon rather than a receding one. Indeed, here the goal is to make the state converge to the terminal set in the required finite time, and not asymptotically as usually guaranteed by a receding horizon strategy;
	\item Second, we adopt a move-blocking strategy (see e.g. \cite{cagienard2007move}) to reduce the computational burden required by the feedback controller. This is motivated by applications like the one described in Section \ref{S:motivation_train}, featuring  values of $k_f$ of the order of several hundreds to thousands. The corresponding number of decision variables, combined with the system nonlinearity, often results in a prohibitive computational complexity when all the predicted control moves are free optimization variables.
\end{itemize}
To the best of our knowledge, the combined presence of nonlinear dynamics, shrinking horizon, and move-blocking strategy is new in the literature. We named the resulting control approach Shrinking horizon Blocking Predictive Control (SBPC). We present next the optimal control problem to be solved at each time step in SBPC, followed by a pseudo-algorithm that realizes this control approach and by a proof of convergence in nominal conditions.

\subsection{Shrinking horizon Blocking Predictive Control (SBPC)}\label{SS:SBPC}
\begin{figure}[h]
\includegraphics[width= \columnwidth]{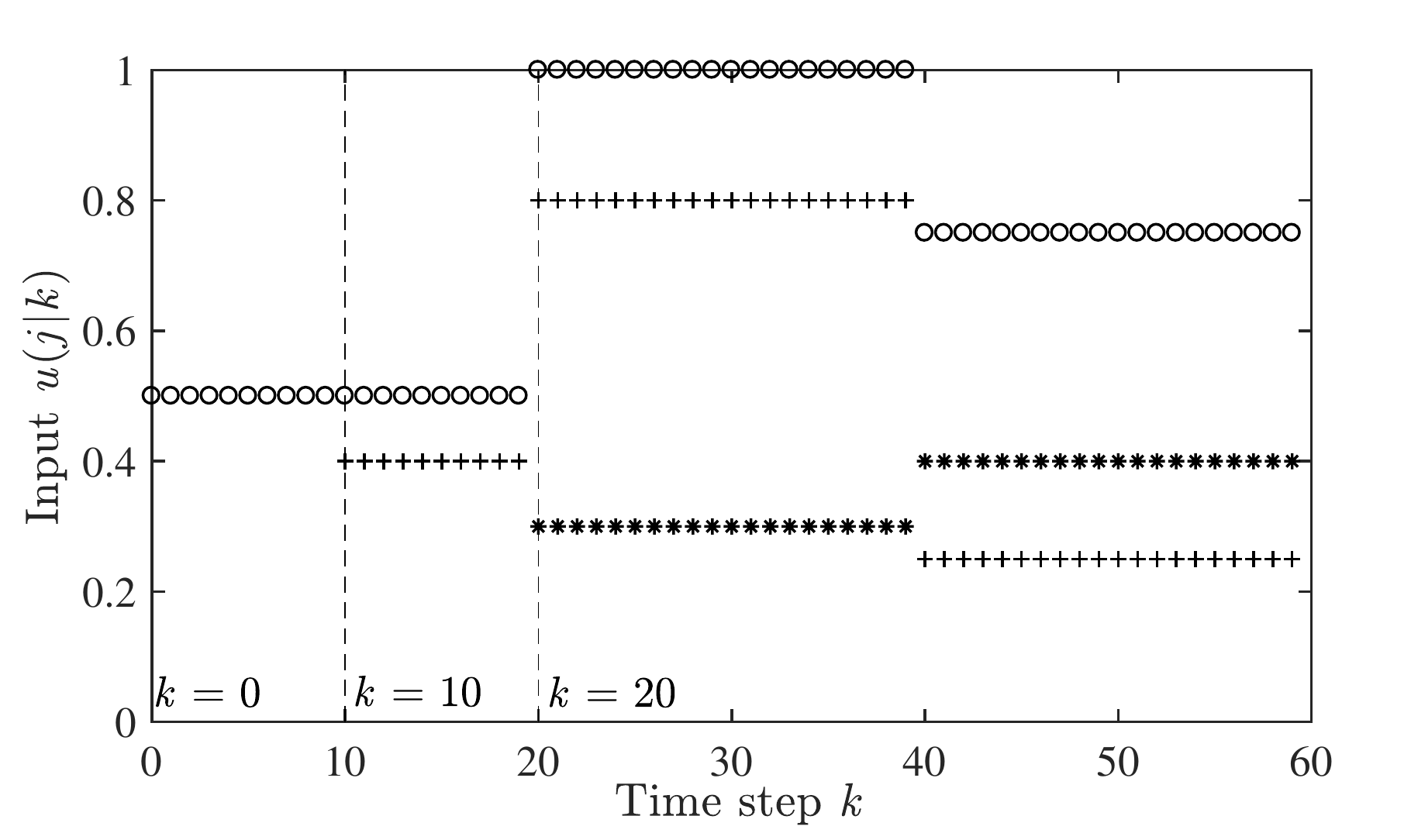}
\caption{SBPC scheme with $L = 20$ and $k_f = 60$. As an example, possible courses of predicted inputs at time $k=0$ (`$\circ$'), $k=10$ (`$+$'), and $k=20$ (`$*$') are depicted. It can be noted that for $k=20$ the number of decision variables reduces from 3 to $N(20)=2$, and that as $k$ increases, the number of blocked moves in the first block decreases.} 
\label{F:sketch_move_blocking}
\end{figure}

We consider a move blocking strategy, where the control input is held constant for a certain time interval. Let us denote with $L$ the maximum number of blocked control moves in each interval within the prediction horizon. Moreover, we consider that each interval contains exactly $L$ blocked moves, except possibly the first one, which can contain a number between 1 and $L$ of blocked input vectors. In this way, for a given value of $k\in[0,\,k_f-1]$, the number $N(k)$ of intervals (i.e. of different blocked input vector values to be optimized) is equal to (see Fig. \ref{F:sketch_move_blocking} for a graphical representation) :
\begin{equation}\label{eq:N_free_inputs}
N(k)= \ceil*{\frac{k_f-k}{L}},
\end{equation}
where $\ceil{.}$ denotes the ceiling operation to the closest integer. Let us denote with $\bm{v}_{N(k)}=\{v(1),\,\ldots,v(N(k))\}\in\mathbb{R}^{m\,N(k)}$, where $v(\cdot)\in\mathbb{U}$, the sequence of free input values to be optimized, i.e. the values that are held constant within each interval of the blocked input sequence (see Fig. \ref{F:sketch_move_blocking}) and with $u(j|k)$ the input vector at time $k+j$ predicted at time $k$. Then, with the described blocking strategy, at each $k$ the values of $u(j|k)$ are computed as:
\begin{equation}\label{eq:u_predict} 
u(j|k)=g(\bm{v}_{N(k)},j,k)\doteq v\left(\floor*{\frac{j+k-\floor*{\frac{k}{L}}L}{L}}+1\right)\\
\end{equation}
Finally, let us denote with $x(j|k),\,j = 0,...,k_f-k$ the state vectors predicted at time $k+j$ starting from the one at time $k$. At each time $k\in[0,\,k_f-1]$, we formulate the following FHOCP:
\begin{subequations}\label{eq:FHOCP_shrinking_nominal} 
	\begin{eqnarray}
	&\min\limits_{\bm{v}_{N(k)}}\;\sum\limits_{j=0}^{k_f-k} \ell(x(j|k),u(j|k))&\\
	&\text{subject to}&\nonumber\\  
	&u(j|k)=g(\bm{v}_{N(k)},j,k),\,j = 0,\ldots,k_f-k-1    &\\                              
	&x(j+1|k) = f(x(j|k),u(j|k)),\,j=0,\ldots,k_f-k-1&\\
	&u(j|k) \in U,\,j=0,\ldots,k_f-k-1&\\
	&x(j|k) \in X,\,j=1,\ldots,k_f-k&\\
	&x(0|k) = x(k)&\\
	&x(k_f-k|k)\in X_f&\label{eq:terminal_constr}
	\end{eqnarray}
\end{subequations}
We denote with $\bm{v}_{N(k)}^*=\{v^*(1),\,\ldots,v^*(N(k))\}$ a solution (in general only locally optimal) of \eqref{eq:FHOCP_shrinking_nominal}. Moreover, we denote with $\bm{x}^*(k)$ and $\bm{u}^*(k)$ the corresponding predicted sequences of state and input vectors:
\begin{subequations}\label{eq:sol_predict_opt} 
\begin{eqnarray}
\bm{x}^*(k)&=&\{x^*(0|k),\ldots, x^*(k_f-k|k)\}\\
\bm{u}^*(k)&=&\{u^*(0|k),\ldots, u^*(k_f-1-k|k)\}\\
\text{where}\nonumber\\
x^*(0|k)&=&x(k)\nonumber\\
x^*(j+1|k)&=&f(x^*(j|k),u^*(j|k))\\
u^*(j|k)&=&g(\bm{v}_{N(k)}^*,j,k)\label{eq:opt_input_nom}
\end{eqnarray}
\end{subequations}
The SBPC strategy is obtained by recursively solving \eqref{eq:FHOCP_shrinking_nominal}, as described by the following pseudo-algorithm.
\begin{algo}\label{alg:SBPC}
	\emph{Nominal SBPC strategy}
	\begin{enumerate}
		\item At sampling instant $k$, measure or estimate the state $x(k)$ and solve the FHOCP  \eqref{eq:FHOCP_shrinking_nominal}. Let $\bm{v}_{N(k)}^*$ be the computed solution;
		\item Apply to the plant the first element of the sequence $\bm{v}_{N(k)}^*$, i.e. the control vector $u(k)=u^*(0|k)=v^*(1)$;
		\item Repeat the procedure from 1) at the next sampling period.\hfill$\square$
	\end{enumerate}
\end{algo}
\noindent Algorithm \ref{alg:SBPC} defines the following feedback control law:
\begin{equation} \label{eq:feedback_SBPC_law}
u(k)=\mu(x(k)):=u^*(0|k),
\end{equation} 
and the resulting model of the closed-loop system is:
\begin{equation} \label{eq:cl_system_SBPC_nominal}
x(k+1)= f(x(k),\mu(x(k))
\end{equation}
\begin{remark}\label{rem:shrinking_SBPC}
 In the approach described so far, the number $N(k)$ of predicted inputs to be optimized decreases from $N(1)=\ceil*{\frac{k_f}{L}}$ to $N(k_f-1)=1$, see Fig. \ref{F:sketch_move_blocking}. Another approach that can be used with little modifications is to keep a constant value of $N=N(1)$, and to reduce the number of blocked input values in each interval as $k$ increases, up until the value $k=k_f-N(1)$ is reached, after which each predicted input vector is a free variable and their number shrinks at each $k$. This second strategy has the advantage to retain more degrees of freedom in the optimization as time approaches its final value.
\end{remark}
We conclude this section with a Lemma on the recursive feasibility of \eqref{eq:FHOCP_shrinking_nominal} and convergence of the state of  \eqref{eq:cl_system_SBPC_nominal} to the terminal set.  
\begin{proposition}\label{pro:nominal_SBPC} Assume that the FHOCP \eqref{eq:FHOCP_shrinking_nominal} is feasible at time $k=0$. Then, the FHOCP \eqref{eq:FHOCP_shrinking_nominal} is recursively feasible at all $k=1,\ldots,k_f-1$ and the state  of the closed loop system \eqref{eq:cl_system_SBPC_nominal} converges to the terminal set $X_f$ at time $k_f$. \hfill$\blacksquare$
\end{proposition}
\textbf{Proof}. Recursive feasibility is established by construction, since at any time $k+1$ one can build a feasible sequence $\bm{v}_{N(k+1)}$ either by taking $\bm{v}_{N(k+1)}=\bm{v}_{N(k)}^*$,  if $N(k+1)=N(k)$, or by taking $\bm{v}_{N(k+1)}=\{v^*(2),\,\ldots,v^*(N(k))\}$ (i.e. the tail of $\bm{v}_{N(k)}^*$), if $N(k+1)=N(k)-1$. Convergence to the terminal set is then achieved by considering that constraint \eqref{eq:terminal_constr} is feasible at time $k=k_f-1$.
\hfill$\blacksquare$\\

\begin{remark}\label{rem:nominal_SBPC} So far, we have disregarded any mismatch between the model $f(x,u)$ and the real plant, like the presence of model uncertainty and external disturbances. For this reason, we termed the SBPC approach of Algorithm \ref{alg:SBPC} the ``nominal'' one. In the next section, we introduce a model of uncertainty, whose form is motivated again by the application described in Section \ref{S:motivation_train}, and two possible variations of Algorithm \ref{alg:SBPC} to deal with it, along with their guaranteed convergence properties. We term these variations the ``relaxed''  approaches, since they  involve the use of suitable soft (i.e. relaxed) constraints to guarantee recursive feasibility.
\end{remark}

\begin{remark}\label{rem:forward_invariance} Convergence to $X_f$ does not necessarily imply forward invariance of such a set under the described control scheme (which is by the way not well defined for $k>k_f$). The capability to keep the state within the terminal set depends on how such a set  is defined (e.g. it holds when $X_f$ contains equilibrium points for the model $f(x,u)$) and in general it is not required by the considered problem setup. This automatically implies that we don't have to assume the existence of any terminal control law as usually done in standard NMPC formulations. On the other hand, in our motivating application the terminal set $X_f$ actually corresponds to an equilibrium point (namely with zero speed, and position equal to the arrival station, see \eqref{eq:train_state_constr}), thus in this case nominal forward invariance is guaranteed for $k>k_f$.
\end{remark}
 
\section{Relaxed SBPC approaches: algorithms and properties}\label{S:relaxed_MPC}

Following Remark \ref{rem:nominal_SBPC}, to model the system uncertainty and disturbances we consider an additive term  $d(k)$ acting on the input vector, i.e.:
\begin{equation} \label{eq:model_disturbance}
\tilde{u}(k)=u(k)+d(k)
\end{equation}
where $\tilde{u}(k)$ is the disturbance-corrupted input provided to the plant. This model represents well all cases where plant uncertainty and exogenous disturbances can be translated into an effect similar to the control input (the so-called matched uncertainty). For example, in our motivating application with straightforward manipulations, equation \eqref{eq:model_disturbance} can describe uncertainty in the train mass, drivetrain specs, track slope and curvature, as well as the discretization of $u^*(0|k)$ and/or misapplication by the human operator in a driver assistance scenario (see Section \ref{S:motivation_train}).\\
\noindent We consider the following assumption on $d$:

\begin{assumption}\label{Ass:bounded disturbance} 
	The disturbance term $d$ belongs to a compact set $\mathbb{D}\subset\mathbb{R}^m$ such that:
	\begin{equation} \label{eq:bound_disturbance}
	\|d\|\leq\overline{d},\,\forall d\in\mathbb{D}
	\end{equation}
	where $\overline{d}\in(0,+\infty)$.\hfill$\square$
\end{assumption}

This assumption holds in many practical cases and in the considered train application as well. We indicate the perturbed state trajectory due to the presence of $d$ as:
\[
\tilde{x}(k+1)=f(\tilde{x}(k),\tilde{u}(k)),\,k=0,\ldots,k_f
\]
where $\tilde{x}(0)=x(0)$. Now, referring to  Proposition \ref{pro:nominal_SBPC}, the convergence guarantees achieved in the nominal case are a direct consequence of the recursive feasibility property, which can be easily lost in presence of the disturbance $d$, due to the deviation of perturbed trajectory from the nominal one. As commonly done in standard NMPC, to retain recursive feasibility, we therefore soften the constraints in the FHOCP. However, in general the use of soft constraints does not guarantee that, in closed-loop operation, the operational constraints are satisfied, or even that the constraint violation is uniformly decreasing as the worst-case disturbance bound $\overline{d}$ gets smaller. For simplicity and to be more specific, from now on let us restrict our analysis to the terminal state constraint in \eqref{eq:FHOCP_terminal_constr}, i.e. $x(k_f)\in X_f$. We do so without loss of generality, since the results and approaches below can be extended to any state constraint in the control problem. On the other hand, in our railway application the terminal state constraint is the most important one from the viewpoint of system performance. The other constraints (velocity limits) are always enforced for safety by modulating traction or by braking.
Let us denote the distance between a point $x$ and a set $X$ as:
\[
\Delta(x,X)=\min\limits_{y\in X}\|x-y\|.
\]
Then, we want to derive a modified SBPC strategy with softened terminal state constraint (to ensure recursive feasibility) that guarantees a property of the following form in closed loop:
\begin{equation} \label{eq:desired_CL_property}
\Delta(\tilde{x}(k_f),X_f)\leq\beta(\overline{d}),\,\beta\in\mathcal{K}.
\end{equation}
That is, the distance between the terminal state and the terminal constraint is bounded by a value that decreases strictly to zero as $\overline{d}\rightarrow 0$. In order to obtain this property, we propose a relaxed SBPC approach using a two-step constraint softening procedure, described next.

\subsection{Two-step relaxed SBPC strategy}

At each time $k$ we consider a strategy consisting of  two optimization problems to be solved in sequence:
\begin{enumerate}
	\item[a)] we compute the best (i.e. smallest) achievable distance between the terminal state and the terminal set, starting from the current perturbed state $\tilde{x}(k)$:\small
\begin{subequations}\label{eq:FHOCP_shrinking_modified1} 
	\begin{eqnarray}
	&\underline{\gamma}=\arg\min\limits_{\bm{v}_{N(k)},\gamma}\; \gamma&\\
	&\text{subject to}&\nonumber\\  
	&u(j|k)=g(\bm{v}_{N(k)},j,k),\,j = 0,\ldots,k_f-k-1    &\\                              
	&x(j+1|k) = f(x(j|k),u(j|k)),\,j=0,\ldots,k_f-k-1&\\
	&u(j|k) \in U,\,j=0,\ldots,k_f-k-1&\\
	&x(j|k) \in X,\,j=1,\ldots,k_f-k&\\
	&x(0|k) = \tilde{x}(k)&\\
	&\Delta(x(k_f-k|k),X_f)\leq\gamma
	\end{eqnarray}
\end{subequations}	\normalsize
\item[b)] we optimize the input sequence using the original cost function, and softening the terminal constraint by $\underline{\gamma}$:\small
\begin{subequations}\label{eq:FHOCP_shrinking_modified2} 
	\begin{eqnarray}
	&\min\limits_{\bm{v}_{N(k)}}\;\sum\limits_{j=0}^{k_f-k} \ell(\tilde{x}(j|k),u(j|k))&\\
	&\text{subject to}&\nonumber\\  
	&u(j|k)=g(\bm{v}_{N(k)},j,k),\,j = 0,\ldots,k_f-k-1    &\\                              
	&x(j+1|k) = f(x(j|k),u(j|k)),\,j=0,\ldots,k_f-k-1&\\
	&u(j|k) \in U,\,j=0,\ldots,k_f-k-1&\\
	&x(j|k) \in X,\,j=1,\ldots,k_f-k&\\
	&x(0|k) = \tilde{x}(k)&\\
	&\Delta(x(k_f-k|k),X_f)\leq\underline{\gamma}&\label{eq:terminal_constr_shrinking_modified2}
	\end{eqnarray}
\end{subequations}\normalsize
\end{enumerate}

By construction, both problems are always feasible (with the caveat that state constraints are considered to be always feasible, as discussed above, otherwise the  softening shall be applied to these constraints as well). We  denote with $\bm{v}_{N(k)}^r$, $\bm{x}^r(k)$ and $\bm{u}^r(k)$ the optimized sequences of decision variables, state and inputs resulting from the solution of \eqref{eq:FHOCP_shrinking_modified2}. The sequences $\bm{x}^r(k)$ and $\bm{u}^r(k)$ are computed from $\bm{v}_{N(k)}^r$ and $\tilde{x}(k)$ as reported in \eqref{eq:sol_predict_opt}. Finally, we note that the disturbance is not explicitly considered in problems \eqref{eq:FHOCP_shrinking_modified1}-\eqref{eq:FHOCP_shrinking_modified2}, which  still employ the nominal model for the predictions.\\
\noindent The resulting relaxed SBPC strategy is implemented by the following pseudo-algorithm.
\begin{algo}\label{alg:relaxed_SBPC}
	\emph{Two-stage relaxed SBPC strategy}
	\begin{enumerate}
		\item At sampling instant $k$, measure or estimate the state $\tilde{x}(k)$ and solve in sequence the optimization problems \eqref{eq:FHOCP_shrinking_modified1}-\eqref{eq:FHOCP_shrinking_modified2}. Let $\bm{v}_{N(k)}^r$ be the computed solution;
		\item Apply to the plant the first element of the sequence $\bm{v}_{N(k)}^r$, i.e. the control vector $u(k)=u^r(0|k)=v^r(1)$;
		\item Repeat the procedure from (1) at the next sampling period.\hfill$\square$
	\end{enumerate}
\end{algo}

\noindent Algorithm \ref{alg:relaxed_SBPC} defines the following feedback control law:
\begin{equation} \label{eq:feedback_relaxed_SBPC_law}
u(k)=\mu^r(\tilde{x}(k)):=u^r(0|k),
\end{equation} 
and the resulting closed-loop dynamics are given by:
\begin{equation} \label{eq:cl_system_relaxed_SBPC}
\tilde{x}(k+1)= f(\tilde{x}(k),\mu^r(\tilde{x}(k))+d(k)).
\end{equation}
\noindent The next result shows that the closed-loop system \eqref{eq:cl_system_relaxed_SBPC} enjoys a uniformly bounded accuracy property of the form \eqref{eq:desired_CL_property}, provided that the nominal SBPC problem \eqref{eq:FHOCP_shrinking_nominal} is feasible at $k=0$.
\begin{theorem}\label{thm:relaxed_SBPC} Let Assumptions \ref{Ass:continuity_f} and \ref{Ass:bounded disturbance} hold and let  the FHOCP \eqref{eq:FHOCP_shrinking_nominal} be feasible at time $k=0$. Then, the terminal state $\tilde{x}(k_f)$ of system \eqref{eq:cl_system_relaxed_SBPC} enjoys  property \eqref{eq:desired_CL_property} with
\begin{equation} \label{eq:desired_CL_property_thm}
\Delta(\tilde{x}(k_f),X_f)\leq\beta(\overline{d})=\sum\limits_{k=0}^{k_f-1}\beta_{k_f-k-1}(\overline{d})
\end{equation}	
where
\begin{equation} \label{eq:desired_CL_property_thm_beta}
\begin{array}{rcl}
\beta_0(\overline{d})&=&a_u(\overline{d})\\
\beta_k(\overline{d})&=&a_u(\overline{d})+a_x(\beta_{k-1}(\overline{d})),\,k=1,\ldots,k_f-1\\
\end{array}
\end{equation}
	 \hfill$\blacksquare$
\end{theorem}
\textbf{Proof}. The proof is by induction. Start at $k=0$ and consider the nominal optimized sequences $\bm{x}^*(k),\,\bm{u}^*(k)$ obtained by solving problem \eqref{eq:FHOCP_shrinking_nominal}. We first evaluate the worst-case perturbation induced by the disturbance with respect to the open-loop state trajectory $\bm{x}^*(k)$. For a sequence of disturbances $d(j|k),\,j=0,\ldots,k_f-k$, the corresponding open-loop input and state trajectories are:
\begin{equation} \label{eq:utilde_pred}
\begin{array}{l}
\tilde{u}(j|k)=u^*(j|k)+d(j|k),\,j=0,\ldots,k_f-k-1\\
\tilde{x}(0|k)=x^*(0|k)\\
\tilde{x}(j+1|k)=f(\tilde{x}(j|k),\tilde{u}(j|k)),\,j=1,\ldots,k_f-k
\end{array}
\end{equation}
From \eqref{eq:continuity_f} we have:
\[
\begin{array}{l}
\|\tilde{x}(1|0)-x^*(1|0)\| =\|f(x(0),\tilde{u}(0|0))-f(x(0),u^*(0|0))\| \leq\\ a_u\left(\|\tilde{u}(0|0)-u^*(0|0)\|\right) \leq a_u\left(\overline{d}\right)=\beta_0\left(\overline{d}\right)
\end{array}
\]
Consider now the perturbation 2-steps ahead: 
\[
\begin{array}{l}
\|\tilde{x}(2|0)-x^*(2|0)\|=\\
\|f(\tilde{x}(1|0),\tilde{u}(1|0))-f(x^*(1|0),u^*(1|0))\|=\\
\|f(\tilde{x}(1|0),\tilde{u}(1|0))-f(\tilde{x}(1|0),u^*(1|0))+\\
f(\tilde{x}(1|0),u^*(1|0))-f(x^*(1|0),u^*(1|0))\|\leq\\
a_u\left(\overline{d}\right)+ a_x\left(\|\tilde{x}(1|0)-x^*(1|0)\|\right)\leq\\ a_u\left(\overline{d}\right)+a_x\left(\beta_0\left(\overline{d}\right)\right)=\beta_1\left(\overline{d}\right).
\end{array}
\]
By iterating up until the second last time step we obtain:
\begin{equation}\label{eq:bound_k_f_0}
\|\tilde{x}(k_f|0)-x^*(k_f|0)\|\leq\beta_{k_f-1}\left(\overline{d}\right),
\end{equation}
where $\beta_{k_f-1}\in\mathcal{K}$ since it is given by compositions and summations of class-$\mathcal{K}$ functions. Since the FHOCP \eqref{eq:FHOCP_shrinking_nominal} is feasible, we  have $x^*(k_f|0)\in X_f$, i.e. $\Delta(x^*(k_f|0),X_f)=0$ and thus:
\begin{equation}\label{eq:beta_0}
\begin{array}{rcl}
\Delta(\tilde{x}(k_f|0),X_f)&\leq&\|\tilde{x}(k_f|0)-x^*(k_f|0)\|+\Delta(x^*(k_f|0),X_f)\\
&\leq&\beta_{k_f-1}(\overline{d}).
\end{array}
\end{equation}
Now consider $k=1$ and the FHOCP \eqref{eq:FHOCP_shrinking_modified1}. If the optimizer is initialized with blocked control moves $\bm{v}_{N(1)}$ such that the tail of the previous optimal sequence $\bm{u}^*(0)$ is applied to the system, the corresponding minimum $\gamma$ in \eqref{eq:FHOCP_shrinking_modified1} results to be upper bounded by $\beta_{k_f-1}(\overline{d})$, in virtue of \eqref{eq:beta_0}. The optimal value $\underline{\gamma}$ is therefore not larger than this bound as well:
\begin{equation}\label{eq:min_dist_k1}
\underline{\gamma}\left|_{k=1}\right.\leq\beta_{k_f-1}(\overline{d}).
\end{equation}
Now take the optimal sequences $\bm{x}^r(1)$ and $\bm{u}^r(1)$ computed by solving the FHOCP \eqref{eq:FHOCP_shrinking_modified2}. By applying the same reasoning as we did for $k=0$, we have (compare with \eqref{eq:bound_k_f_0}):
\begin{equation}\label{eq:beta_1}
\|\tilde{x}(k_f|1)-x^r(k_f|1)\|\leq\beta_{k_f-2}(\overline{d}).
\end{equation}
Moreover, equation \eqref{eq:min_dist_k1} implies that the solution of \eqref{eq:FHOCP_shrinking_modified2} satisfies the following inequality:
\begin{equation}\label{eq:dist_1}
\Delta(x^r(k_f|1),X_f)\leq\underline{\gamma}\left|_{k=1}\right.
\end{equation}
From \eqref{eq:min_dist_k1}-\eqref{eq:dist_1} we have:
\[
\begin{array}{l}
\Delta(\tilde{x}(k_f|1),X_f)\leq\|\tilde{x}(k_f|1)-x^r(k_f|1)\|+\Delta(x^r(k_f|1),X_f)\leq\\
\beta_{k_f-2}(\overline{d})+\beta_{k_f-1}(\overline{d})=\sum\limits_{k=0}^{1}\beta_{k_f-k-1}(\overline{d}).
\end{array}
\]
By applying recursively the same arguments, the bound \eqref{eq:desired_CL_property_thm} is obtained.	 \hfill$\blacksquare$\\
\noindent Theorem \ref{thm:relaxed_SBPC} indicates that the worst-case distance between the terminal state and the terminal set is bounded by a value which is zero for $\overline{d}=0$ and increases strictly with the disturbance bound. In the considered railway application this means that, for example, the worst-case accuracy degradation in reaching the terminal station due to a discretization of the input, as done in the driver assistance mode, is proportional to the largest employed quantization interval of the input handle. This result provides a theoretical justification to the proposed two-step relaxed SBPC approach. The bound \eqref{eq:desired_CL_property_thm} is conservative, since it essentially results from the accumulation of worst-case perturbations induced by the disturbance on the open-loop trajectories computed at each $k$. As we show in our simulation results, in practice the resulting closed-loop performance are usually very close to those of the nominal case, thanks to recursive optimization in the feedback control loop.

\subsection{Multi-objective relaxed SBPC strategy}
As an alternative to the two-step approach described above, one can also consider a multi-objective minimization:
\begin{subequations}\label{eq:FHOCP_shrinking_multiobjective} 
	\begin{eqnarray}
	&\min\limits_{\bm{v}_{N(k)},\beta}\;\sum\limits_{j=0}^{k_f-k} \ell(\tilde{x}(j|k),u(j|k))+\omega\gamma &\\
	&\text{subject to}&\nonumber\\  
	&u(j|k)=g(\bm{v}_{N(k)},j,k),\,j = 0,\ldots,k_f-k-1    &\\                              
	&x(j+1|k) = f(\tilde{x}(j|k),u(j|k)),\,j=0,\ldots,k_f-k-1&\\
	&u(j|k) \in U,\,j=0,\ldots,k_f-k-1&\\
	&x(j|k) \in X,\,j=1,\ldots,k_f-k&\\
	&x(0|k) = \tilde{x}(k)&\\
	&\Delta(x(k_f-k|k),X_f)\leq\gamma&
	\end{eqnarray}
\end{subequations}
where $\omega$ is a positive weight on the scalar $\gamma$. Problem \eqref{eq:FHOCP_shrinking_multiobjective} can be solved in Algorithm \eqref{alg:relaxed_SBPC} in place of problems \eqref{eq:FHOCP_shrinking_modified1}-\eqref{eq:FHOCP_shrinking_modified2}. In this case, the advantage is that a trade-off between constraint relaxation and performance can be set by tuning $\omega$. Regarding the guaranteed bounds on constraint violation, with arguments similar to those employed in \cite{fagianoTeel2013} one can show that, at each $k\in[0,k_f-1]$, for any $\varepsilon>0$ there exists a finite value of $\omega$ such that the distance between the terminal state and the terminal set is smaller than $\gamma_{k_f-k-1}(\overline{d})+\varepsilon$. Thus, with large-enough $\omega$, one can recover the behavior obtained with the two-step relaxed SBPC approach. The theoretical derivation is omitted for the sake of brevity, as it is a rather minor extension of the results of  \cite{fagianoTeel2013}.

\end{document}